\documentclass[fleqn,usenatbib]{mnras}
\usepackage{newtxtext,newtxmath}
\usepackage[T1]{fontenc}
\usepackage{mathptmx} 
\usepackage{graphicx} 
\usepackage{amsmath}  
\usepackage{amssymb}  

\newcommand{\rsun}{{\rm R}_\odot}
\newcommand{\msun}{{\rm M}_\odot}
\newcommand{\msunyr}{{\rm M}_\odot\,{\rm yr}^{-1}}
\newcommand{\nh}{n_{\rm H}}
\newcommand{\mh}{m_{\rm H}}
\newcommand{\cc}{{\rm cm^{-3}}}
\newcommand{\kms}{{\rm km\,s^{-1}}}
\newcommand{\vrad}{v_{\rm rad}}
\newcommand{\vrot}{v_{\rm rot}}
\newcommand{\vturb}{v_{\rm turb}}
\newcommand{\cs}{c_{\rm s}}
\newcommand{\ralfven}{R_{\rm A}}

\title[Rotation of first stars]
{Angular momentum transfer in primordial discs and the rotation of the first stars}

\author[S. Hirano and V. Bromm]{
Shingo Hirano\thanks{E-mail: shirano@astro.as.utexas.edu}
and
Volker Bromm
\\
Department of Astronomy, University of Texas at Austin, Austin, TX 78712, USA
}

\date{Accepted ---. Received ---; in original form ---}
\pubyear{2018}

\begin{document}

\label{firstpage}
\pagerange{\pageref{firstpage}--\pageref{lastpage}}
\maketitle

\begin{abstract}
We investigate the rotation velocity of the first stars by modelling the angular momentum transfer in the primordial accretion disc.
Assessing the impact of magnetic braking, we consider the transition in angular momentum transport mode at the Alfv$\acute{\rm e}$n radius, from the dynamically dominated free-fall accretion to the magnetically dominated solid-body one.
The accreting protostar at the centre of the primordial star-forming cloud rotates with close to breakup speed in the case without magnetic fields.
Considering a physically-motivated model for small-scale turbulent dynamo amplification, we find that stellar rotation speed quickly declines if a large fraction of the initial turbulent energy is converted to magnetic energy ($\gtrsim 0.14$).
Alternatively, if the dynamo process were inefficient, for amplification due to flux-freezing, stars would become slow rotators if the pre-galactic magnetic field strength is above a critical value, $\simeq\!10^{-8.2}$\,G, evaluated at a scale of $\nh = 1\,\cc$, which is significantly higher than plausible cosmological seed values ($\sim\!10^{-15}$\,G).
Because of the rapid decline of the stellar rotational speed over a narrow range in model parameters, the first stars encounter a bimodal fate: rapid rotation at almost the breakup level, or the near absence of any rotation.
\end{abstract}


\section{Introduction}
\label{sec:intro}

One of the outstanding challenges in modern cosmology is the formation process of the first generation of stars, the so-called Population III (Pop~III).
They influence all subsequent star and galaxy evolution in the early Universe, through their input of ionizing radiation and heavy chemical elements. 
The latter sensitively depends on the final fates of Pop~III stars \citep{karlsson13}.
There have been no direct observations yet, but the nature of the first stars has been elucidated by theoretical studies, in particular with numerical simulations of increasing physical realism
\citep[for recent reviews, see][]{bromm13,greif15,barkana16}.
Furthermore, there are a number of indirect empirical constraints, exhibiting the imprint of the first stars. 
Among them are extremely metal-poor (second-generation) stars, which retain the characteristic fingerprint of their parent Pop~III stars in the chemical abundance pattern \citep[e.g.][]{frebel05,beers05,keller14},
metal-poor stellar systems in high-$z$ galaxies, such as the enigmatic CR7 source \citep{sobral15}, and 
the gravitational wave signal emitted from the merger of massive black hole binaries \citep[e.g.][]{GW150914,hartwig16}.
There is need to further develop the theoretical model for the formation and evolution of the first stars to predict their observational signature, in light of the upcoming suite of next-generation telescopes, such as the {\it James Webb Space Telescope (JWST)}.

The evolution and death of Pop~III stars mainly depend on two parameters, their mass and rotation state.
Specifically, stellar mass is the crucial parameter in determining their overall impact on the early Universe
\citep[e.g.][]{heger02,heger03}.
According to theory, non-rotating zero-metallicity stars end their lives as core-collapse supernovae (CCSNe) in the mass range $15 < M_*/\msun < 40$ (on the main sequence), pair-instability supernovae (PISNe) in the range $140 < M_*/\msun < 260$, or directly collapse into black holes (BHs) for $40 < M_*/\msun < 140$ or $260 < M_*/\msun$.
Recent three-dimensional, gravito-radiation-hydrodynamics simulations examine in detail how massive the first stars can grow during the accretion phase \citep[e.g.][]{hosokawa11,hosokawa16,stacy12,stacy16,susa13,susa14,hirano14}.

As the second governing parameter, stellar rotation has been considered \citep[e.g.][]{maeder12}.
The dependence of stellar evolution on rotation is stronger for lower-metallicity \citep{brott11}, emphasizing the need to extend our understanding to Pop~III.
More rapid rotation enhances both mass loss and internal chemical mixing, which in turn greatly affect the path of stellar evolution, e.g. by establishing homogeneous chemical abundance distributions.
Select consequences are the nucleosynthetic abundance pattern in massive Pop~III stars \citep[e.g.][]{chiappini11}, and the lowering of the minimum mass for a PISN to occur \citep[e.g.][]{chatzopoulos12,yoon12}. 
The latter is crucial to correctly assess the prospects for future PISN surveys with {\it JWST} \citep[e.g.][]{hummel12}.
Furthermore, extremely rapid rotators with more than half the breakup speed encounter violent fates, such as gamma-ray bursts (GRBs), providing highly energetic probes of the early Universe \citep[e.g.][]{bromm06,levan16,toma16}. 
As a final example, very massive ($\ge\!250\,\msun$), rapidly rotating, low-metallicity stars might constitute single progenitors for a subset of the observed gravitational wave signals \citep{dorazio17}.

However, the degree of stellar rotation in Pop~III stars remains highly uncertain.
One has to contend with the computational difficulty of following the protostellar accretion process for long times ($\sim$Myr), while also achieving the extremely high densities to resolve the protostellar surface.
The most precise three-dimensional hydrodynamics simulation for first star growth could only follow the accretion for ten years after initial protostar formation \citep{greif12}.
Other studies considered longer accretion times, but needed to evaluate the accreted angular momentum at the numerical resolution scale, which was much larger than the actual stellar surface \citep{stacy11rot,stacy13rot}.
The currently available results suggest a final rotation close to the breakup speed, but have not yet taken into account the entire angular momentum transfer over all relevant scales.

To develop a more realistic picture, the possible spin-down from such near-breakup rotation due to magnetic stresses should be considered.
This question is related to the so-called `angular momentum problem' in present-day (Population~I) star formation, where the observed initial angular momentum of a cloud core must be efficiently redistributed or removed during collapse \citep[e.g.][]{mckee07,larson10}.
As a possible solution, the angular momentum endowed to the protostar via the accreting gas may be limited by the magnetic braking on the accretion disc.
Specifically, there are two main theoretical ideas to achieve this regulation \citep[for a review, see][]{littlefair14}: magnetic star-disc interaction \citep[`disc locking'; e.g.][]{matt05mnras,matt10} and stellar winds \citep[e.g.][]{matt05apj,matt12}.
In a recent synthesis, \cite{rosen12} develop an angular momentum evolution model for massive stars, considering both magnetic and gravitational torques.

In existing simulations of primordial accretion discs, however, magnetic effects are often neglected because of the tiny cosmological seed strength \citep{xu08,widrow12}.
In some numerical studies, the fields are significantly amplified during the subsequent cloud collapse \citep{sur10,sur12,federrath11,turk12}.
The resulting magnetic stresses could thus influence the long-term disc evolution, accretion and fragmentation process \citep{machida13}.
Later on, a magnetically-driven jet could affect the subsequent evolution \citep[e.g.][]{latif16}.
However, the overall impact on stellar rotation remains uncertain.
In this paper, we construct a simple, semi-analytical model to evaluate angular momentum transport, mediated by the primordial accretion disc.
We derive a critical magnetic field strength above which the stellar rotation quickly declines via magnetic braking.
This idealized estimate needs to be updated with future, fully self-consistent simulations, but is useful as an initial exploration.

The remainder of the paper is organized as follows.
We begin by describing our semi-numerical methodology in Section~\ref{sec:methods}.
Section~\ref{sec:res} presents the model results, in particular the rotational speed of the first stars depending on models of the magnetic field strength amplification.
In Section~\ref{sec:dis}, we provide our main conclusions and discuss their implications.

\section{Methodology}
\label{sec:methods}

\begin{figure}
\includegraphics[width=\columnwidth]{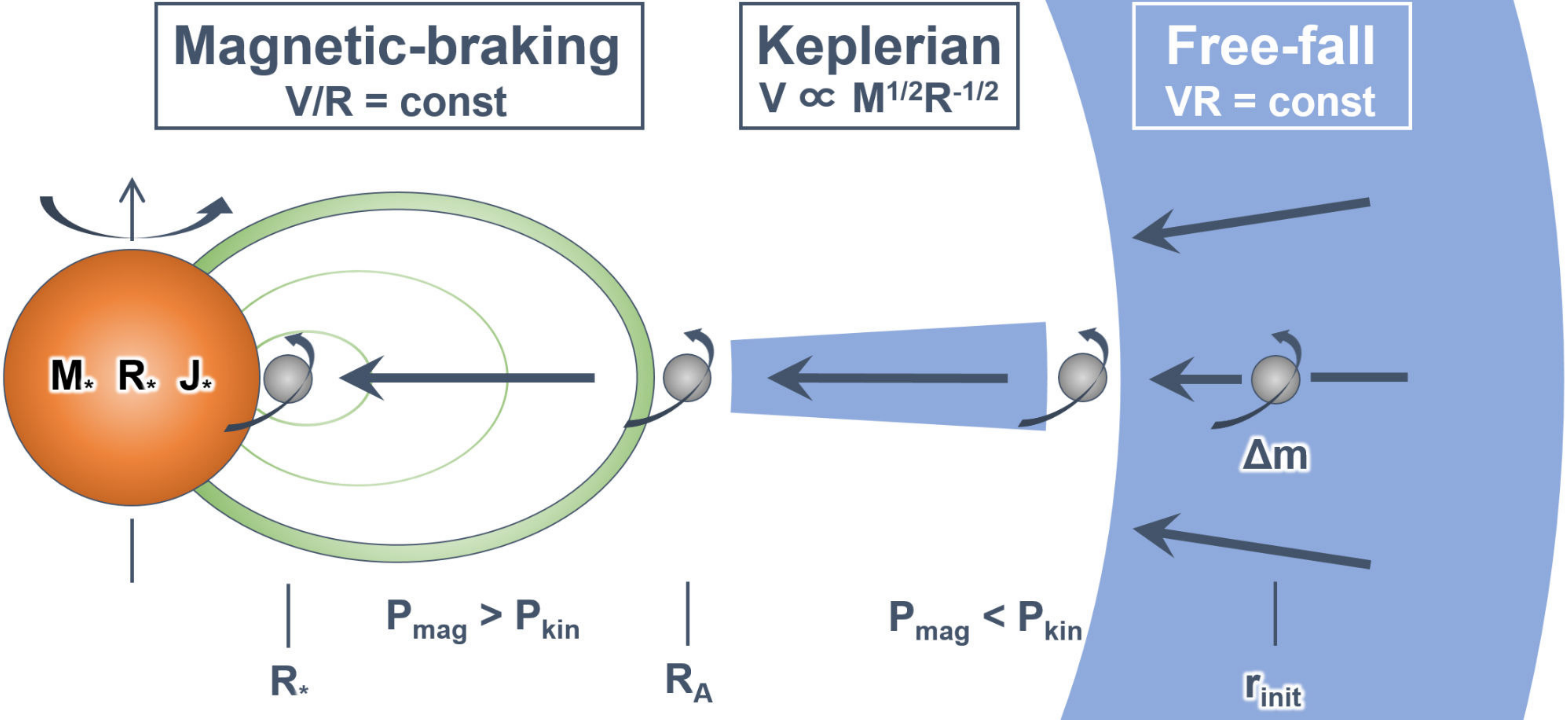}
\caption{
Schematic overview of the angular momentum transfer model. 
The accreted gas experiences a sequence of transport modes, from free-fall farther out in the collapsing envelope, to viscous torques in a Keplerian outer disc, to the inner disc, where magnetic stresses dominate ($P_{\rm mag}>P_{\rm kin} $). 
The transition occurs at $\ralfven$, the Alfv$\acute{\rm e}$n radius. 
In the first regime, specific angular momentum is conserved, and in the latter, solid-body rotation obtains.
}
\label{fig1}
\end{figure}

The aim of this study is to evaluate the dependence of stellar rotation on magnetic field strength for Pop~III stars.
To circumvent the difficulty of rigorous magneto-hydrodynamic (MHD) numerical simulations, we construct an idealized semi-analytical model (see Fig.~\ref{fig1}).
We compute the transfer of angular momentum onto the protostar via accreting gas using the results of cosmological hydrodynamical simulations.
Although these baseline simulations do not include any MHD effects, we still have a valid representation of the gas density and kinematics during the early stages of collapse, when magnetic fields are not yet dynamically important (see Fig.~\ref{fig2}).
We then analytically estimate the distribution of magnetic field strength (Section~\ref{sec:methods_mag}).
The mode of angular momentum transfer undergoes a transition at the Alfv$\acute{\rm e}$n radius, $\ralfven$, inside of which magnetic pressure overcomes the dynamical ram pressure (Section~\ref{sec:methods_angmom}).
Our model evaluates the resulting angular momentum imparted to the protostar surface (Section~\ref{sec:methods_radius}), and yields the resulting stellar rotational velocity (Section~\ref{sec:methods_protostar}).

\subsection{Field amplification during cloud collapse}
\label{sec:methods_mag}

The distribution of magnetic field strength throughout the star-forming cloud is a necessary ingredient for our modelling, but is not self-consistently calculated within the cosmological simulation.
This is justified in the cosmological context due to the initial weakness of any seed field, such as those generated through astrophysical mechanisms, including the Biermann battery \citep[e.g.][]{biermann50} and Weibel instability \citep[e.g.][]{schlickeiser03}.
During the subsequent cloud collapse, however, the $B$-field can be further amplified via gravitational compression \citep[e.g.][]{xu08}, the small-scale turbulent dynamo process, whose amplification rate depends on numerical resolution \citep[e.g.][]{sur10,sur12,federrath11,turk12,schober12}, and the magneto-rotational instability \citep[MRI; e.g.][]{balbus91,hawley91,balbus98,silk06}.
During the later accretion phase, the tangled magnetic fields resulting from the small-scale dynamo could be re-aligned coherently via the $\alpha$-$\Omega$ dynamo in differentially rotating discs \citep[e.g.][]{pudritz89,tan04,latif16}.
The magnetic field strength may finally reach the equipartition value, providing dynamically important magnetic stresses.

To date, there have been no MHD simulations which can investigate the entire amplification process of magnetic field strength, from the tiny cosmological seed to the final value inside a newborn star.
Therefore, we adopt idealized amplification models during the collapse stage of the star-forming cloud and determine the magnetic field distribution at the beginning of the accretion phase.

\subsubsection{Flux-freezing}
\label{sec:methods_mag_ff}

\begin{figure}
\includegraphics[width=\columnwidth]{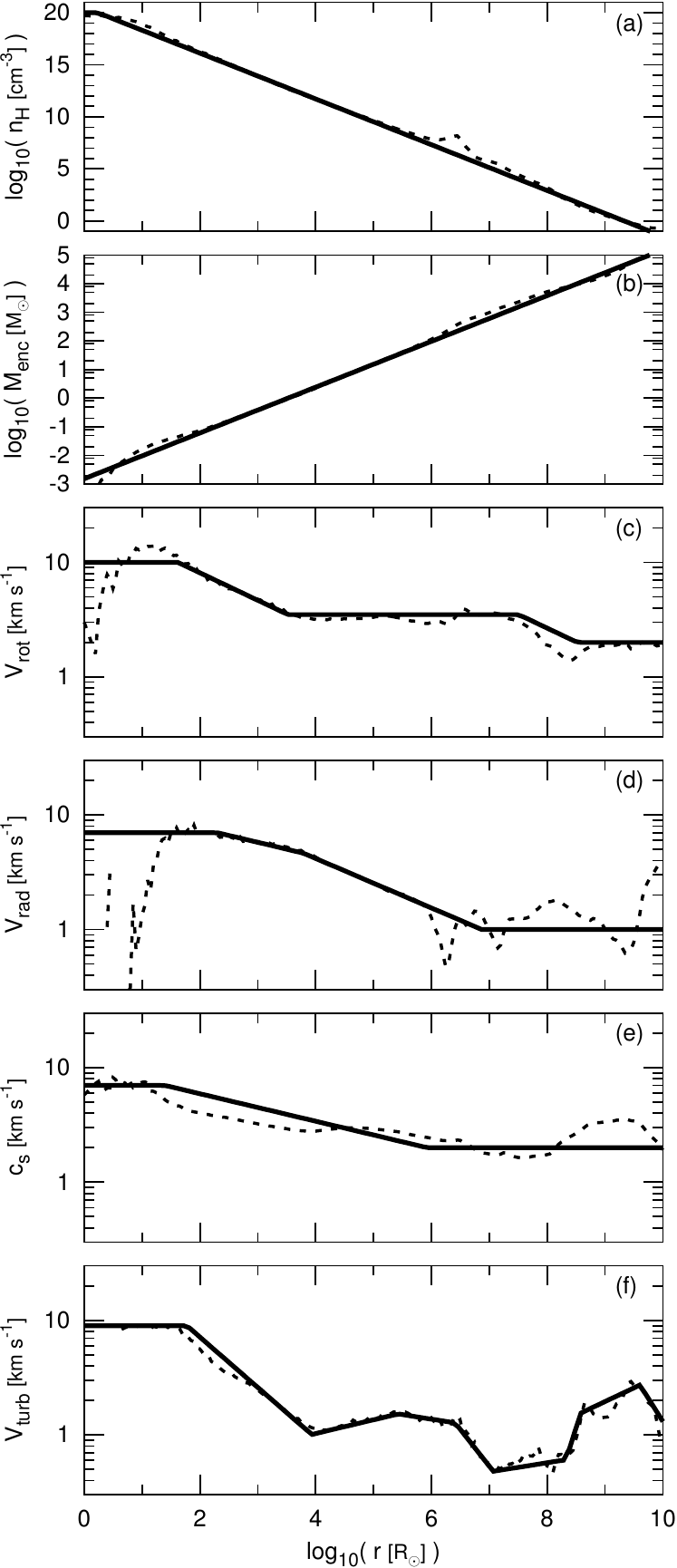}
\caption{
Initial conditions for semi-analytic $B$-field model. 
We show radial distributions of gas properties in the collapsed primordial cloud, as obtained in the underlying cosmological simulation:
({\it a}) gas number density,
({\it b}) enclosed gas mass,
({\it c}) rotation velocity,
({\it d}) radial infall velocity,
({\it e}) sound speed, and
({\it f}) turbulent velocity.
The solid lines show analytical fits, provided in Appendix~\ref{app:fitfunc}, to the simulation results (dashed lines), evaluated at the beginning of the protostellar accretion phase.
}
\label{fig2}
\end{figure}

\begin{figure}
\includegraphics[width=0.95\columnwidth]{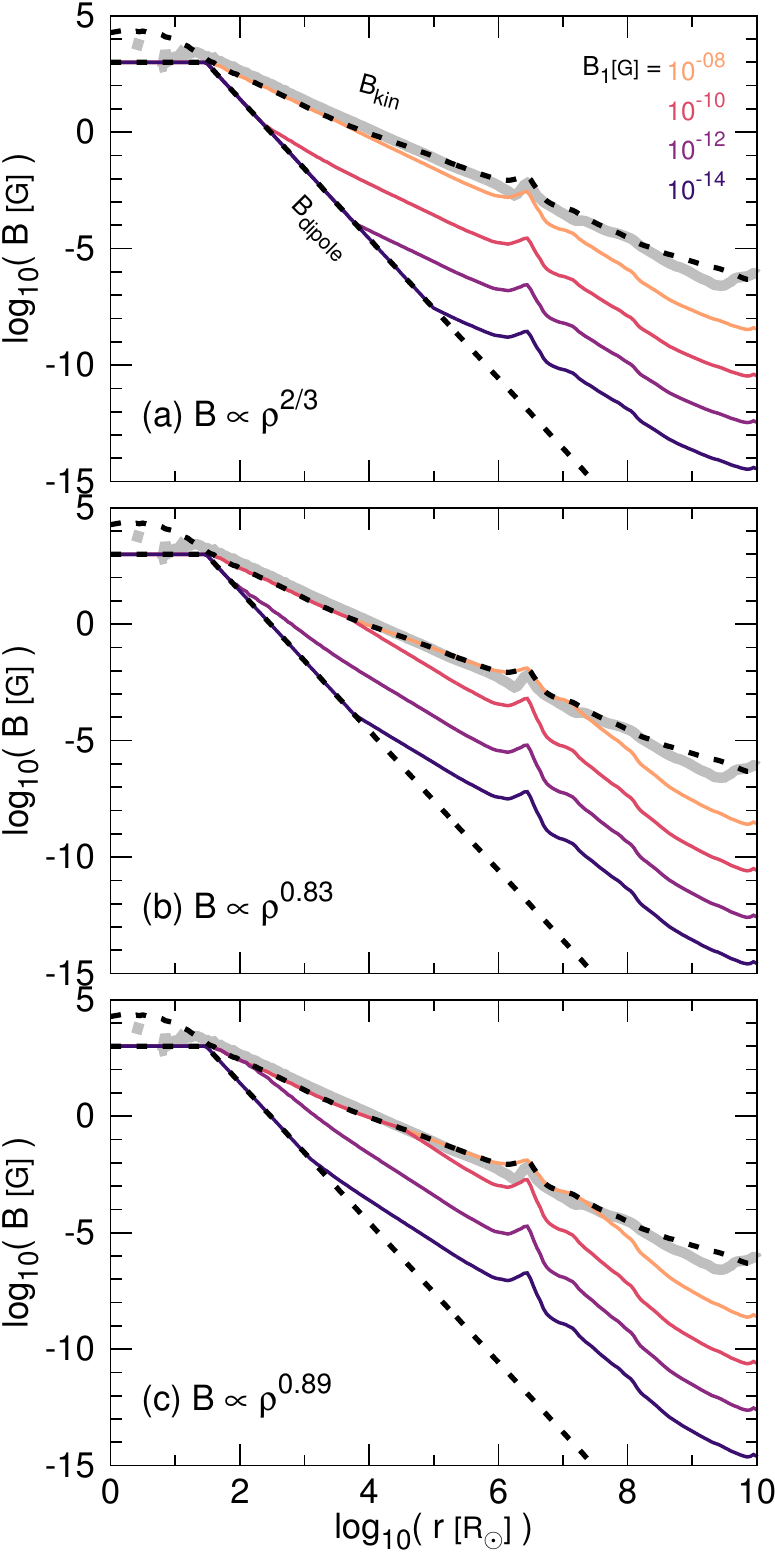}
\caption{
Radial distributions of magnetic field strength, amplified during the collapse of a primordial star-forming cloud: ({\it a}) $B = B_1 \nh^{2/3}$, ({\it b}) $B = B_1 \nh^{0.83}$, and ({\it c}) $B = B_1 \nh^{0.89}$.
The line colours represent the initial $B$-field strength when $\nh = 1\,\cc$, $B_1 = 10^{-14}$, $10^{-12}$, $10^{-10}$ and $10^{-8}$\,G in panels ({\it a} - {\it c}), respectively.
The two dashed lines are the maximum and minimum field strengths, $B_{\rm kin}$ and $B_{\rm dipole}$, respectively (see Section~\ref{sec:methods_mag_range} for further discussion).
The transition in angular momentum transfer mode occurs where the amplified field strength reaches the thick grey line, representing the condition $B = (4 \piup \rho \vrad^2)^{1/2}$.
}
\label{fig3}
\end{figure}

The small magnetic seed field can be further amplified during cloud compression.
If the $B$-field is tightly coupled to the collapsing gas through flux-freezing, the resulting amplification is described by a power law, $B \propto \nh^{2/3}$, where $\nh\,(= \rho/\mh)$ is the gas number density, normalized to the proton mass $\mh$.
As basic model parameter, we specify the cosmological seed field, $B_1$, before the cloud collapse begins, at a gas number density of $\nh\ = 1\,\cc$.
Figure~\ref{fig3}(a) shows the resulting field amplification, employing the simulation data from Fig.~\ref{fig2}.

\begin{figure}
\includegraphics[width=0.95\columnwidth]{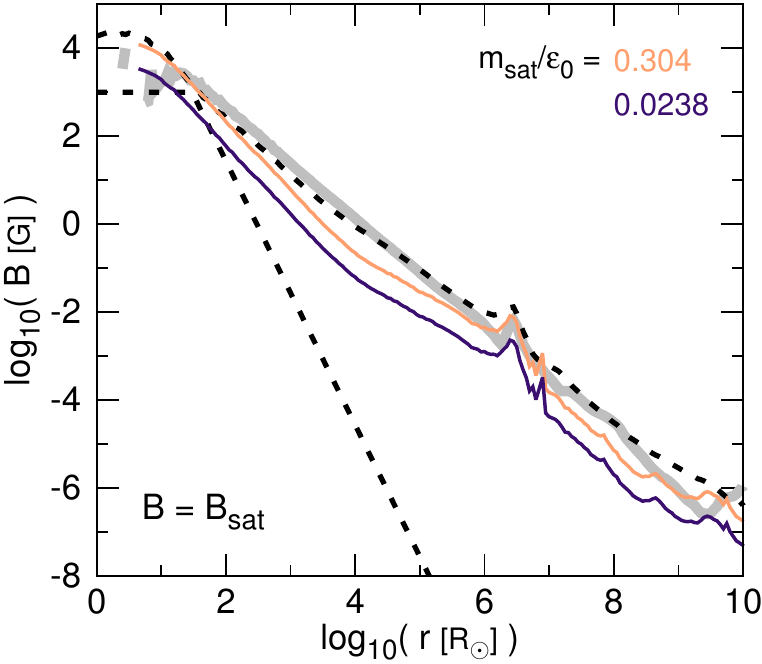}
\caption{
Magnetic field strength vs. radius, similar to Fig.~\ref{fig3}, but for the physically-motivated dynamo model, such that $B = B_{\rm sat}$.
Here, line colours represent the saturation level, $m_{\rm sat}/\epsilon_0 = 0.304$ and $0.0238$.
}
\label{fig4}
\end{figure}

\subsubsection{Small-scale turbulent dynamo}
\label{sec:methods_mag_dynamo}

The turbulent dynamo can lead to further amplification.
To exemplify this, cosmological MHD simulations provide power-law fits to their results, to be compared to the flux-freezing expression, specifically $B \propto \nh^{0.83}$ \citep[][]{federrath11} and $B \propto \nh^{0.89}$ \citep[][]{turk12}, shown in Figs~\ref{fig3}(b) and (c).
However, the amplification level via the turbulent small-scale dynamo depends on the numerical resolution, and no MHD simulation has reached convergence.
Following a different route, \cite{schober15} present a physically-motivated model for the small-scale dynamo, indicating that the magnetic field can be highly amplified by energy conversion from the initial turbulent kinetic energy.
Depending on the detailed properties of the turbulence, their study derives an upper limit of $\sim\!30$\% for the resulting saturation level.

Here, we apply the maximum saturation level of the magnetic field strength obtained from the physically-motivated model \citep{schober15}, $\alpha_{\rm sat} = m_{\rm sat} / \epsilon_0$, where $m_{\rm sat}$ is the magnetic energy density at saturation and $\epsilon_0$ the initial turbulent kinetic energy.
The maximally amplified magnetic field strength, $B_{\rm sat}$, then becomes
\begin{equation}
\frac{B_{\rm sat}(R)^2}{8 \pi} = \alpha_{\rm sat} \frac{\rho v_{\rm turb}(R)^2}{2} \, ,
\label{eq:Bsat}
\end{equation}
where $v_{\rm turb}$ is the turbulent velocity, depicted in Fig.~\ref{fig2}(f).
The saturation level, $\alpha_{\rm sat}$, depends on the microphysical properties of the turbulence, and is determined by the magnetic Prandtl number, ${\rm Pm} = {\rm Rm}/{\rm Re}$, where ${\rm Rm}$ and ${\rm Re}$ are the magnetic and hydrodynamical Reynolds numbers, respectively.
Assuming Kolmogorov turbulence for the Pop~III protostellar disc \citep{stacy16}, we show the resulting field amplification in Fig.~\ref{fig4}, for two extreme cases: $\alpha_{\rm sat} = 0.304$ (${\rm Pm} \gg 1$) and $0.0238$ (${\rm Pm} \ll 1$).

\subsubsection{Lower and upper limits}
\label{sec:methods_mag_range}

We also consider lower and upper limits for the $B$-field distribution.
As minimum value, we assume a dipole field anchored to the central protostar.
The perpendicular component in the equatorial plane at a distance $r$ from the star is
\begin{equation}
B_{\rm dipole} = B_* \left( \frac{r}{R_*} \right)^{-3} \, ,
\label{eq:Beq}
\end{equation}
where $B_*$ is the field strength at the stellar surface $R_*$. 
We adopt $B_* = 1$\,kG as fiducial value, based on the surface field strength observed for present-day, pre-main-sequence stars \citep[e.g.][]{johns-krull07}. Massive Pop~III protostars may well behave differently, but we defer such a more complete exploration to future work.
For the radius of the accreting protostar, for simplicity we adopt a constant value $R_* = 30\,\rsun$, consistent with the radii employed in our protostellar evolution model, discussed below (see Fig.~\ref{fig5}c).

We estimate the upper limit for the $B$-field by assuming that the magnetic pressure, or energy density, must be less than the local dynamical pressure to continue cloud collapse and gas accretion, $B^2 \leq 4 \piup \rho \cs^2$, where $\cs$ is the sound speed.
The resulting maximum $B$-field is
\begin{equation}
B_{\rm kin} = 0.14\,{\rm G}\,\left( \frac{\nh}{10^{10}\,\cc} \right)^{1/2} \left( \frac{\cs}{3\,\kms} \right) \, .
\label{eq:Bmaxkin}
\end{equation}
Our results are largely consistent with recent studies of the dynamo-amplified field strength in primordial discs, arguing that fields can reach close to the saturation, or energy equipartition, level \citep{latif16}.\footnote{Note that our estimate for the maximally amplified field, $B_{\rm kin}$, is consistent with the upper limit for the saturation level achievable for the small-scale dynamo \citep{schober15}.}
We finally assume that a significant fraction of the magnetic stress can be organized into a large-scale, coherent topology, e.g. by following the small-scale turbulent dynamo with an $\alpha$-$\Omega$ dynamo, or by imposing the stellar dipole field on to the surrounding disc.
As a consequence, the inner disc, where magnetic stresses may dominate, could be forced into solid-body rotation (see the discussion below).

\subsection{Angular momentum transport modes}
\label{sec:methods_angmom}

Our model considers the magnetic impact on angular momentum transfer in the infalling gaseous material.
We assume that the rotational velocity evolution undergoes a transition at the Alfv$\acute{\rm e}$n radius, $\ralfven$ (see Fig.~\ref{fig1}).
At $r > \ralfven$, where magnetic pressure is negligible, the gas accretion proceeds dynamically. 
Initially, the gas experiences free-fall collapse with constant angular momentum, such that $\vrot r = \text{\rm const}$, where $\vrot$ is the rotational velocity. 
After the formation of a centrifugally supported disc, the gas eventually joins the disc and migrates through the accretion disc with the Keplerian rotation speed, $v_{\rm Kep} = \sqrt{G M_{\rm enc}(r) / r}$, where $M_{\rm enc}(r)\gtrsim M_{\ast}$ is the mass enclosed within radius $r$, which is in turn close to the mass of the central protostar, $M_{\ast}$.
At $r < \ralfven$, on the other hand, magnetic stresses dominate the accretion process. 
We further assume that, as a consequence of the dominant magnetic stress, the disc is locked into solid-body rotation, such that $\vrot/r = \text{\rm const}$.

The angular momentum evolution of accreting gas thus depends on the initial distance from the star, as follows.
For a gas element accreted from an initial radius, $r_{\rm init}$, the rotational velocity at the stellar surface, $R_*$, is
\begin{equation}
  \vrot(R_*) = \left( \frac{R_*}{r_{\rm init}} \right) \vrot(r_{\rm init}) \, ,
  \label{eq:V*_MHD}
\end{equation}
if $r_{\rm init} \le \ralfven$, and
\begin{equation}
  \vrot(R_*) = \left( \frac{R_*}{\ralfven} \right) \vrot(\ralfven) = \left( \frac{R_*}{\ralfven} \right) \min \left[ \left( \frac{r_{\rm init}}{\ralfven} \right) \vrot(r_{\rm init}), v_{\rm Kep} \right] \, ,
\label{eq:V*_MHD+FF}
\end{equation}
if $r_{\rm init} > \ralfven$, respectively. In the latter expression, we evaluate the Keplerian speed at the Alvf$\acute{\rm e}$n radius, $v_{\rm Kep} = v_{\rm Kep}(\ralfven)$.
The character of dynamical infall switches from free-fall collapse to disc accretion when the rotational velocity reaches $v_{\rm Kep}$.

The mode transition occurs where the magnetic pressure balances the ram pressure of spherical (free-fall) accretion, $B^2 = 4 \piup \rho \vrad^2$, where $\vrad$ is the radial infall velocity.
Here, the dynamical and magnetic cloud properties are given by their respective radial distributions (Figs.~\ref{fig2}, \ref{fig3} and \ref{fig4}).
This criterion thus implicitly defines the Alfv$\acute{\rm e}$n radius.
It is convenient to work in terms of a local function, $\ralfven(r)$, to evaluate the transition criterion, as follows.
Substituting the shell-crossing mass accretion rate, $\rho = \dot{M} / 4 \piup R^2 \vrad$, we arrive at the expression
\begin{equation}
\ralfven(r) \simeq 9 \times 10^{4}\,\rsun \left( \frac{B}{0.1\,{\rm G}} \right)^{-1} \left( \frac{\dot{M}}{0.04\,\msunyr} \right)^{1/2} \left( \frac{\vrad}{3\,\kms} \right)^{1/2} \, .
\label{eq:Ralfven}
\end{equation}
The Alfv$\acute{\rm e}$n radius is then given by the condition $\ralfven(r)=r$.
During the accretion process, each gas element migrates inwards toward the central protostar, possibly entering a region where magnetic pressure becomes dominant.
Because the radial distribution of the infall velocity is very similar to that of the sound speed (see Figs.~\ref{fig2}d and e), the crossing points between the $B$-field distribution and the upper limit, $B_{\rm kin}$, represent the transition scales, depending on the seed field strength (Fig.~\ref{fig3}) and the saturation level (Fig.~\ref{fig4}).

\subsection{Protostellar radial evolution}
\label{sec:methods_radius}

We model the radial evolution of the accreting protostar as in \cite{stacy11rot}, and refer the reader to their section~3.2.2 for details.
A newborn protostellar core gradually expands during the initial adiabatic accretion phase, growing as
\begin{equation}
R_1 =  50 \left( \frac{M_*}{1\,\msun} \right)^{1/3} \left( \frac{\dot{M}_*}{\dot{M}_{\rm crit}} \right)^{1/3}\,\rsun \, , \\
\label{eq:Rstar1}
\end{equation}
where $\dot{M}_{\rm crit} = 4.4 \times 10^{-3}\,\msunyr$ is the critical accretion rate above which stellar radiative feedback, opposing continued gas accretion, becomes efficient \citep{omukai2003}.
The stellar radius shrinks during the following Kelvin-Helmholtz (KH) contraction phase as
\begin{equation}
R_2 = 140 \left( \frac{M_*}{10\,\msun} \right)^{-2} \left( \frac{\dot{M}_*}{\dot{M}_{\rm crit}} \right)\,\rsun \, . \\
\label{eq:Rstar2}
\end{equation}
The phase shift occurs when $R_2$ becomes smaller than $R_1$.
If the stellar rotational speed reaches the breakup value during the KH contraction, we assume that the star rotates at the breakup speed by slowing the radial contraction to
\begin{equation}
R_3 = \exp \left[ \ln\,R_{\rm *,pre} - \frac{{\rm d}}{{\rm d}t}\ln\,M_{*} \right]\,\rsun \, , \\
\label{eq:Rstar3}
\end{equation}
where $R_{\rm *,pre}$ denotes the stellar radius at the previous step.
Finally, the contracting star reaches the zero-age main sequence (ZAMS) at a radius
\begin{equation}
R_4 = 4.65 \left( \frac{M_*}{100\,\msun} \right)^{0.61}\,\rsun \, . \\
\label{eq:Rstar4}
\end{equation}
Recently, this model has been updated \citep{stacy16}, but we adopt the earlier version for simplicity.

\subsection{Protostellar angular momentum}
\label{sec:methods_protostar}

Employing the protostellar radial evolution model above, we proceed to evaluate the angular momentum history of the accreting protostar as follows.
The time-dependent mass growth, $M_*(t)$, is obtained from the simulation data, represented by an analytical fit $M_*(t)$ (Fig.~\ref{fig5}a and Equation~\ref{eq:Mstar}).
When the stellar mass grows from $M_*(t-{\rm d}t)$ to $M_*(t)$, the newly accreted gas mass is ${\rm d}M_* = M_*(t) - M_*(t-{\rm d}t)$.
We assume that the gas element migrates from a distance $r$ where $M_{\rm enc}(r) = M_*$, arriving at the protostar with a rotational velocity given by Equations~\ref{eq:V*_MHD} or \ref{eq:V*_MHD+FF}.
Then, the total angular momentum acquired by the star is $J_*(t) = \int_0^t \dot{J}_{\rm acc}(t) {\rm d}t$, where $\dot{J}_{\rm acc} = \vrot(R_*) R_* {\rm d}M_*/{\rm d}t$.
The resulting angular momentum of the star with mass $M_*$ and radius $R_*$, modelled as a solidly rotating sphere for simplicity, is
$J_* = \Omega_* I_*$, where
$\Omega_*$ is the angular velocity and $I_* = 2 M_* R_*^2/5$ the moment of inertia.
Finally, the equatorial rotational velocity of the accreting star is
\begin{equation}
V_* = R_* \Omega_* = \frac{5 J_*}{2 R_* M_*} \, .
\label{eq:V_*}
\end{equation}
We note that we ignore the possible inclination of the accretion flow with respect to the rotational axis
\citep[e.g.][]{bate10,fielding15}, and assume accretion through an equatorial disc.

\subsection{Time evolution of accreting protostar}
\label{sec:methods_data}

\begin{figure}
\includegraphics[width=\columnwidth]{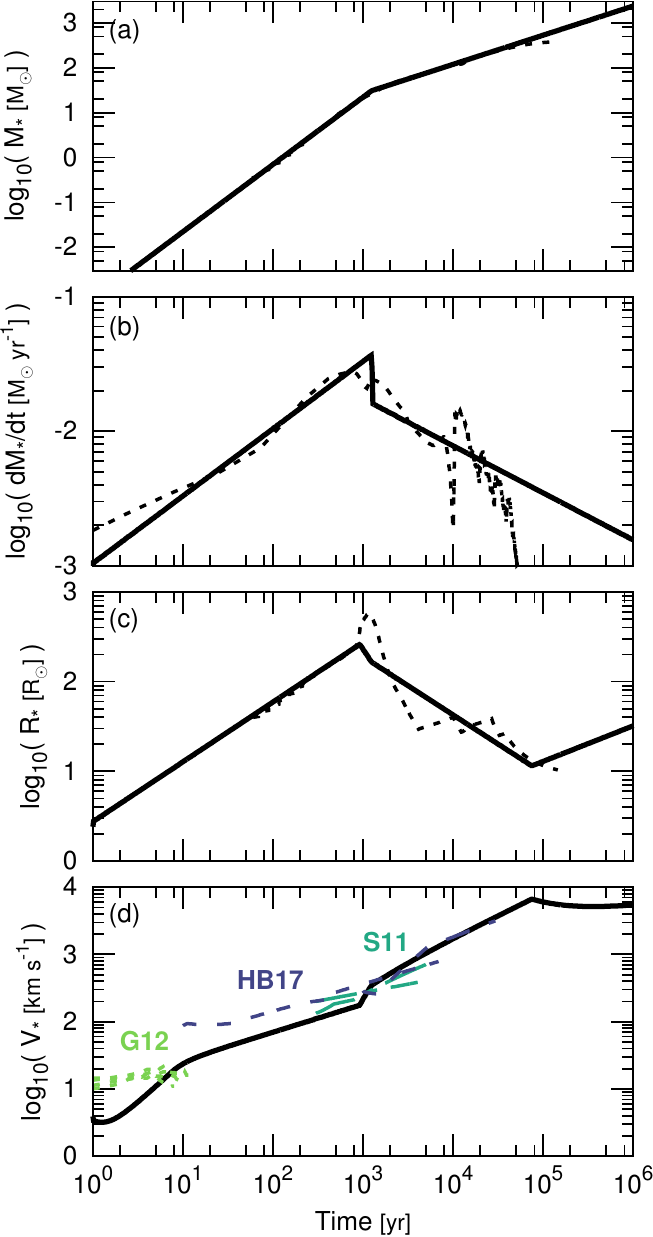}
\caption{
Semi-analytical model of accreting protostar, 
formed at the centre of a primordial cloud (see Fig.~\ref{fig2}). We show key stellar parameters, as a function of time since initial protostar formation:
({\it a}) stellar mass,
({\it b}) accretion rate,
({\it c}) stellar radius, and 
({\it d}) rotational velocity. 
The dashed lines in the top three panels show the original data obtained by performing a two-dimensional radiation-hydrodynamic simulation \citep{hirano14}. 
The solid lines in panels ({\it a}) and ({\it b}) are fitting functions to the simulation data (see Appendix~\ref{app:fitfunc}), whereas in panel ({\it c}), the solid line represents the approximate model for the radial evolution from Section~\ref{sec:methods_radius}.
Finally, the solid line in the bottom panel depicts the rotation history of the Pop~III star in the absence of magnetic fields, as calculated with our idealized model.
For comparison, we also show results from other simulations discussed in more detail in Appendix~\ref{app:previous}: G12 \citep[][dotted lines]{greif12,stacy13rot}, HB17 \citep[][dashed]{hirano17a}, and S11 \citep[][dot-dashed]{stacy11rot}.
It is evident that our zero-field result is largely consistent with those simulations. 
}
\label{fig5}
\end{figure}

Figure~\ref{fig5} summarize the evolution of the Pop~III protostar, resulting from our idealized modelling.
The analytical fitting functions well reproduce the simulation results (Figs.~\ref{fig5}a and b).
Although the fitting formulae smooth over some detailed features, such as the episodic accretion, which can act to quickly expand the stellar radius \citep{hosokawa16,sakurai16}, the overall evolution is properly represented.
We note that the extrapolation to times beyond $\sim\!10^5$\,yr is very uncertain.

Figure~\ref{fig5}(d) shows the stellar rotational velocity evolution without magnetic fields present.
This is the baseline case where we can compare to results from previous simulations of Pop~III star formation.
Specifically, we consider a set of simulations with different spatial resolutions \citep[][see also Appendix~\ref{app:previous}]{greif12,stacy11rot,hirano17a}.
Each simulation provides estimates for the mass ($M_{\rm c}$), radius ($R_{\rm c}$) and angular momentum ($j_{\rm c}$) of the central core, at the resolved scales (see Fig.~\ref{figA1}).
The lowest resolution simulation covers the entire history of the protostellar accretion phase ($\sim\!10^5$\,yr), whereas at highest-resolution, only the initial $\sim\!10$\,yr are covered.
For simplicity, we assume that the mass accretion histories at the protostellar surface and the resolved scale are the same, such that $M_*(t) = M_{\rm c}(t)$.
Applying our idealized model (see above) to this input data, we derive estimates for the resulting stellar rotation velocity ({\it coloured lines} in Fig.~\ref{fig5}).

These cases represent an intermediate approach between fully consistent, ab-initio simulations  
of protostellar core formation, and our idealized, semi-analytic formalism here.
Encouragingly, the extrapolated rotational velocities at different radial and time scales from the simulations are well reproduced by our model.
The future goal, however, is to achieve realistic simulations, capable of resolving the protostellar core scale, while at the same time being able to continue to the end of the accretion phase.
As argued in this paper, it will be crucial to also include MHD effects in such next-generation simulations.

\begin{figure*}
\begin{tabular}{cc}
\includegraphics[width=\columnwidth]{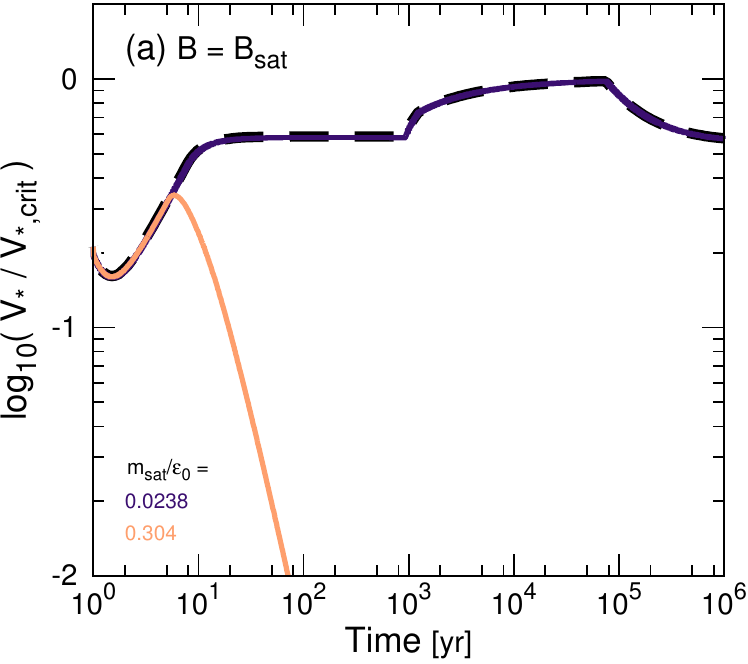} &
\includegraphics[width=\columnwidth]{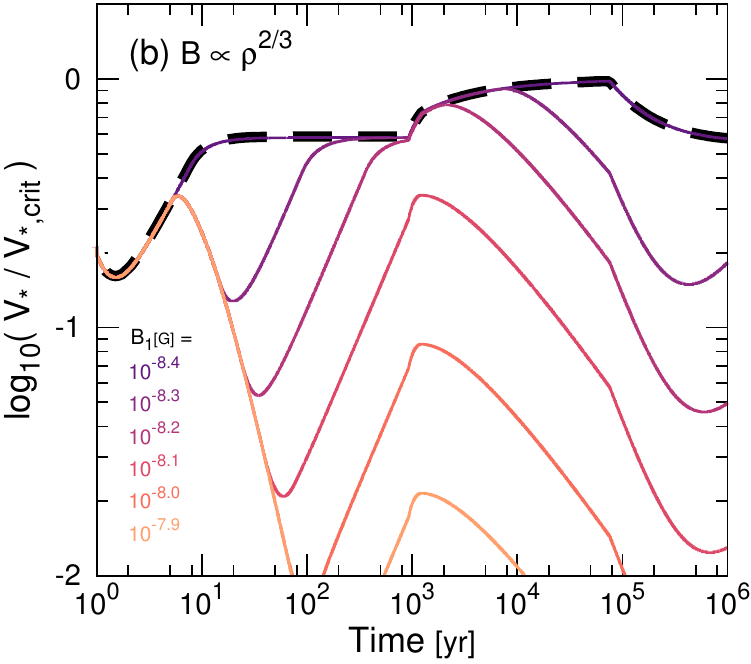}
\end{tabular}
\caption{
Rotation velocity normalized to the breakup speed of an accreting Pop~III protostar, adopting two models for magnetic field amplification: 
({\it a}) small-scale dynamo, where $B = B_{\rm sat}$ (Equation~\ref{eq:Bsat}), and
({\it b}) flux-freezing, where $B = B_1 (\nh/1\,\cc)^{2/3}$.
For the physically-motivated dynamo case, we in turn adopt two extreme cases, $\alpha_{\rm sat} = m_{\rm sat}/\epsilon_0 = 0.304$ and $0.0238$.
For the flux-freezing one, we explore a grid of large-scale seed values, $B_1 = 10^{-8.4}$--$10^{-7.9}$\,G, chosen to bracket the critical field strength, above which magnetic braking becomes efficient.
The dashed lines represent the reference history of rotation velocity in the absence of magnetic fields.
}
\label{fig6}
\end{figure*}

\section{Stellar rotation histories}
\label{sec:res}

We investigate how the stellar rotation speed changes due to the magnetic braking by considering the switch in angular momentum transfer mode at $\ralfven$ (Equation~\ref{eq:Ralfven}).
In so doing, we explore different amplification models for the magnetic field in the primordial star-forming cloud.

In the absence of any magnetic field, $B_1 = 0$, the stellar rotational velocity gradually grows to $\sim\!100\,{\rm km\,s^{-1}}$ during the first expansion phase, and further increases in the KH contraction stage (see Fig.~\ref{fig5}d).
In this case, the rotation speed continues to increase up to $\gtrsim\!1$,$000\,{\rm km\,s^{-1}}$ due to the angular momentum introduced via the accreting gas, and finally reaches an asymptotic value at the ZAMS transition, $\sim\!10^5$\,yr after initial protostar formation.
Due to the final spin-up, the rotational velocity almost reaches the breakup speed, $V_{\rm *,crit}$, towards the end of the stellar mass growth (dashed line in Fig.~\ref{fig6}).
Without any effects capable of removing angular momentum, massive Pop~III stars would therefore typically become rapid rotators, as has previously been established \citep[e.g.][]{stacy11rot,stacy13rot}.

The coloured lines in Fig.~\ref{fig6} represent stellar rotation histories, normalized to the breakup speed, $V_{\rm *,crit}$, for different amplification models.
Magnetic braking suppresses the stellar spin-up within $\ralfven$, such that
the rotational degree decreases during the accretion of material within $\ralfven$, endowed with a reduced rotational velocity of $\vrot(R_*) / \vrot(r) < 1$ (Equation~\ref{eq:V*_MHD}). The subsequent evolution sensitively depends on the efficiency of
the small-scale dynamo process, to be discussed next.

\subsection{Efficient dynamo}
\label{sec:res_dynamo}

In Figure~\ref{fig6}(a), we show results for the physically-motivated model of the small-scale turbulent dynamo.
The stellar rotation histories are markedly different depending on the assumed saturation levels, measured by $\alpha_{\rm sat}$.
In the maximally amplified case with $\alpha_{\rm sat} = 0.304$ (${\rm Pm} \gg 1$), the transition at the Alfv$\acute{\rm e}$n radius occurs at larger disc radius (Fig.~\ref{fig4}), such that the rotational degree continues to decrease for a prolonged time.
On the other hand, when $\alpha_{\rm sat} = 0.0238$ (${\rm Pm} \ll 1$), the transition occurs close to the stellar surface, such that there is no effective magnetic braking.
When the saturation level falls below $\alpha_{\rm sat} \simeq 0.14$, the stellar rotation history abruptly changes from the non-rotating case to the one at almost breakup speed. Therefore, if the saturation level of the small-scale dynamo amplification exceeds this critical value, Pop III stars become slow rotators, independent of the initial magnetic field.

\subsection{Inefficient dynamo}
\label{sec:res_nodynamo}

In the case of an inefficient small-scale turbulent dynamo ($\alpha_{\rm sat} < 0.14$), some degree of effective magnetic braking could still occur, if compressional flux-freezing were to sufficiently amplify the magnetic seed field. Any suppression of the stellar spin-up would then depend on the pre-galactic magnetic field strength, $B_1$ (Fig.~\ref{fig6}b). The resulting rotation histories are now more complex. After the
initial braking phase, the rotation velocity increases again, when the gas outside $\ralfven$ migrates to the stellar surface, with rotational velocity depending on the initial radius as $\vrot(R_*) / \vrot(r) = R_* r / \ralfven^2$ (Equation~\ref{eq:V*_MHD+FF}).
During the KH contraction phase, the rotation speed also decreases because of the dependence of accreting angular momentum, ${\rm d}J_{\rm acc} = \vrot(R_*) R_* {\rm d}M_* = (R_* / \ralfven)^2 \vrot(r) r {\rm d}M_*$, on the contracting stellar radius $R_*$.
The final rise at $t \sim 10^5$\,yr occurs after the ZAMS is reached.

There is a drastic decline in final rotation velocity, $V_* / V_{\rm *,crit} \simeq 0.2$ at $t = 10^5$\,yr, for $B_1 \simeq 10^{-8.2}$\,G.
If the pre-galactic seed field is less than this threshold value, the rotation recovers the high spin of the zero $B$-field case.
Because this decline occurs over such a narrow range in $B_1$, the first stars encounter a bimodal fate, where they either rapidly rotate, with close to the breakup speed, or exhibit virtually no rotation at all.
Simulated cosmological seed fields, $\sim\!10^{-15}$\,G at $\nh = 1\,\cc$ \citep{xu08}, are less than this critical value, such that the first stars are predicted to typically be rapid rotators unless there are additional amplification processes present.

\section{Summary and Conclusions}
\label{sec:dis}

Overall, our model suggests that the spin-down of the first stars requires efficient small-scale dynamo activity, or flux-freezing amplification from relatively high initial magnetic field strengths. Specifically, our analysis indicates that there is a bifurcation in the final rotational state of a Pop~III star, exhibiting rotation either at near-zero or close to breakup speed. We may speculate that such bimodality is also
reflected in observables, such as the nucleosynthetic patterns in extremely metal-poor stars \citep[e.g.][]{chiappini11}. 
One key parameter in deciding which rotational fate is more common, or typical, is the magnetic Prandtl number in the disc. Existing studies suggest that this number varies over many orders of magnitude during the collapse \citep[e.g.][]{schober12}, and guidance from present-day star formation is inconclusive \citep[e.g.][]{oishi2011}. 
Therefore, future MHD simulations are needed to self-consistently study the properties of discs in primordial star formation.

The final degree of rotation is crucially important for the subsequent stellar evolution of Pop~III stars.
A key example is rotationally induced mixing, reducing the chemical abundance gradient throughout the star \citep[e.g.][]{maeder12}.
Specifically, massive stars with close to breakup rotation, $V_* / V_{\rm *,crit} \gtrsim 0.5$, can undergo chemically homogeneous evolution, and enter a Wolf-Rayet phase \citep[e.g.][for the primordial case]{yoon05,woosley06}.
The final fate and death of the first stars also sensitively depend on the rotation velocity \citep[e.g.][see their fig.~12]{yoon12}.
Rapid stellar rotation will decrease the mass threshold for triggering energetic events, such as PISNe and hypernovae \citep[e.g.][]{nomoto03,chatzopoulos12}, or facilitate the occurrence of collapsar-driven GRBs at high redshifts \citep[e.g.][]{wang12}.
As an important implication for Pop~III feedback, rapidly rotating stars are less puffed up and thus bluer, resulting in the increased emission of ionizing photons.
The surrounding H${\rm \, II}$ region would then be more extended, thereby affecting the impact of subsequent SN feedback.

The detailed differences in stellar evolution sensitively depend on the stellar model employed \citep[e.g.][]{ekstrom08}.
In addition, magnetic fields may also affect stellar evolution.
An example is the Spruit-Taylor dynamo which can transport angular momentum from the core to the envelope \citep{spruit02}.
There are a number of rotationally-induced dynamical effects, such as a bar-like instability, which enables angular momentum transfer via gravitational torques, acting to slow down the core \citep{lin11}.
Recently, \cite{lee16} report that rapid rotation can control the mass growth of massive Pop~III stars, through the so-called $\Omega \Gamma$ limit, where centrifugal forces boost the effectiveness of radiation pressure.
As another example, \cite{haemmerle17} present an evolution model of a supermassive Pop~III star which implies that the surface rotation velocity is maintained at less than $10$--$20$\% of the Keplerian value to satisfy the $\Omega \Gamma$ limit.

The magnetic braking can also change the morphology of the accreting gas envelope and disc.
\cite{machida13} perform magneto-hydrodynamic simulations to investigate disc formation and fragmentation in the magnetized primordial cloud, whose magnetic field is amplified according to $B \propto \rho^{2/3}$.
They conclude that the star formation mode changes from binary or multiple fragmentation to single protostar formation, if the ambient $B$-field exceeds $10^{-12}$\,--\,$10^{-13} (\nh/1\,\cc)$\,G.
This critical value is less than the threshold one in our model, $\simeq\!\!10^{-8.2}$\,G for the same, flux-freezing, amplification case.
Thus, disc fragmentation may be easier to affect by magnetic fields than the rotation state of a Pop~III star.
The efficiency of angular momentum transport will affect the shape of the accreting envelope, thus modulating how strongly photo-evaporative feedback can impact the protostellar assembly process \citep[e.g.][]{mckee08,hosokawa16}.

The current study assesses how the magnetic braking of accreting gas regulates the resulting stellar rotation of the first stars.
We present an idealized study to explore the basic physics, and in so doing neglect a number of effects, such as possible outflows, which can remove stellar angular momentum \citep[e.g.][]{littlefair14}.
Ultimately, we will have to consider the neglected mechanisms by performing improved simulations, but our simple model evaluates the key effect on stellar rotation, covering the entire accretion process, which at present is not yet amenable to fully ab-initio simulations.
It is intriguing that the emergence of the first stars at the end of the cosmic dark ages may be so intimately coupled to primordial magneto-genesis.

\section*{Acknowledgements}

We would like to thank the anonymous referee for carefully reading and comments which improve the quality of this paper.
This work was supported by Grant-in-Aid for JSPS Overseas Research Fellowships to SH, and by NSF grant AST-1413501 to VB.

\bibliographystyle{mnras}
\bibliography{biblio}

\appendix

\section{Simulation data and fitting functions}
\label{app:fitfunc}

The numerical data used in our modelling (dashed lines in Figs.~\ref{fig2} and \ref{fig5}) are obtained from a three-dimensional cosmological simulation \citep{hirano14}.
The simulation set-up and methodology are briefly summarized in section~3.1 in \cite{hirano17a}, where the same cloud is adopted as the initial condition.
We represent the simulation data with the following fitting functions, spherically averaged around the new-born protostar, to be used in our model.
The gas number density is (Fig.~\ref{fig2}a)
\begin{equation}
\nh(r) = 10^{10}\,(r/6.06 \times 10^4\,\rsun)^{-2.2}\,\cc \, ,
\label{eq:fit_nH}
\end{equation}
the enclosed gas mass (Fig.~\ref{fig2}b)
\begin{equation}
M_{\rm enc}(r) = \int_0^r 4 \piup r^2 \rho {\rm d}r = 100\,(r/1.05 \times 10^6\,\rsun)\,\msun \, ,
\label{eq:fit_Menc}
\end{equation}
the rotation velocity, as a function of enclosed mass (Fig.~\ref{fig2}c), 
\begin{equation}
\vrot(M_{\rm enc}) = 
\begin{cases}
10.0 \, , \\
3.5\,(M_{\rm enc}/1\,\msun)^{-0.3} \, , \\
3.5 \, , \\
2.0\,(M_{\rm enc}/10^4\,\msun)^{-0.3} \, , \\
2.0\,\kms \, , \\
\end{cases}
\label{eq:fit_vrot}
\end{equation}
the radial velocity (Fig.~\ref{fig2}d) 
\begin{equation}
\vrad(M_{\rm enc}) = 
\begin{cases}
7.0 \, , \\
7.0\,(M_{\rm enc}/0.106\,\msun)^{-0.15} \, , \\
1.0\,(M_{\rm enc}/481.4\,\msun)^{-0.27} \, , \\ 
1.0\,\kms \, , \\
\end{cases}
\label{eq:fit_vrad}
\end{equation}
the sound speed (Fig.~\ref{fig2}e)
\begin{equation}
\cs(M_{\rm enc}) = 
\begin{cases}
7.0 \, , \\
7.0 \,(M_{\rm enc}/0.02\,\msun)^{-0.15} \, , \\
2.0\,\kms \, , \\
\end{cases}
\label{eq:fit_cs}
\end{equation}
and the turbulent velocity (Fig.~\ref{fig2}f)
\begin{equation}
\vturb(M_{\rm enc}) = 
\begin{cases}
9.0 \, , \\
9.0 \,(M_{\rm enc}/0.04\,\msun)^{-0.55} \, , \\
1.0 \,(M_{\rm enc}/2\,\msun)^{-0.15} \, , \\
1.5 \,(M_{\rm enc}/40\,\msun)^{-0.1} \, , \\
1.3 \,(M_{\rm enc}/200\,\msun)^{-0.8} \, , \\
0.5 \,(M_{\rm enc}/10^3\,\msun)^{0.1} \, , \\
0.8 \,(M_{\rm enc}/8 \times 10^3\,\msun)^{2} \, , \\
1.5 \,(M_{\rm enc}/10^4\,\msun)^{0.3} \, , \\
4.0 \,(M_{\rm enc}/5 \times 10^4\,\msun)^{-1} \,\kms \, . \\
\end{cases}
\label{eq:fit_vturb}
\end{equation}
The mass accretion history of the protostar (Fig.~\ref{fig5}a) is also fitted as
\begin{equation}
M_*(t) = 
\begin{cases}
30.92\,(t/1250\,{\rm yr})^{1.50} \, , \\
30.92\,(t/1250\,{\rm yr})^{0.65}\,\msun \, . \\
\end{cases}
\label{eq:Mstar}
\end{equation}

\section{Summary of previous simulations}
\label{app:previous}

\begin{figure}
\includegraphics[width=\columnwidth]{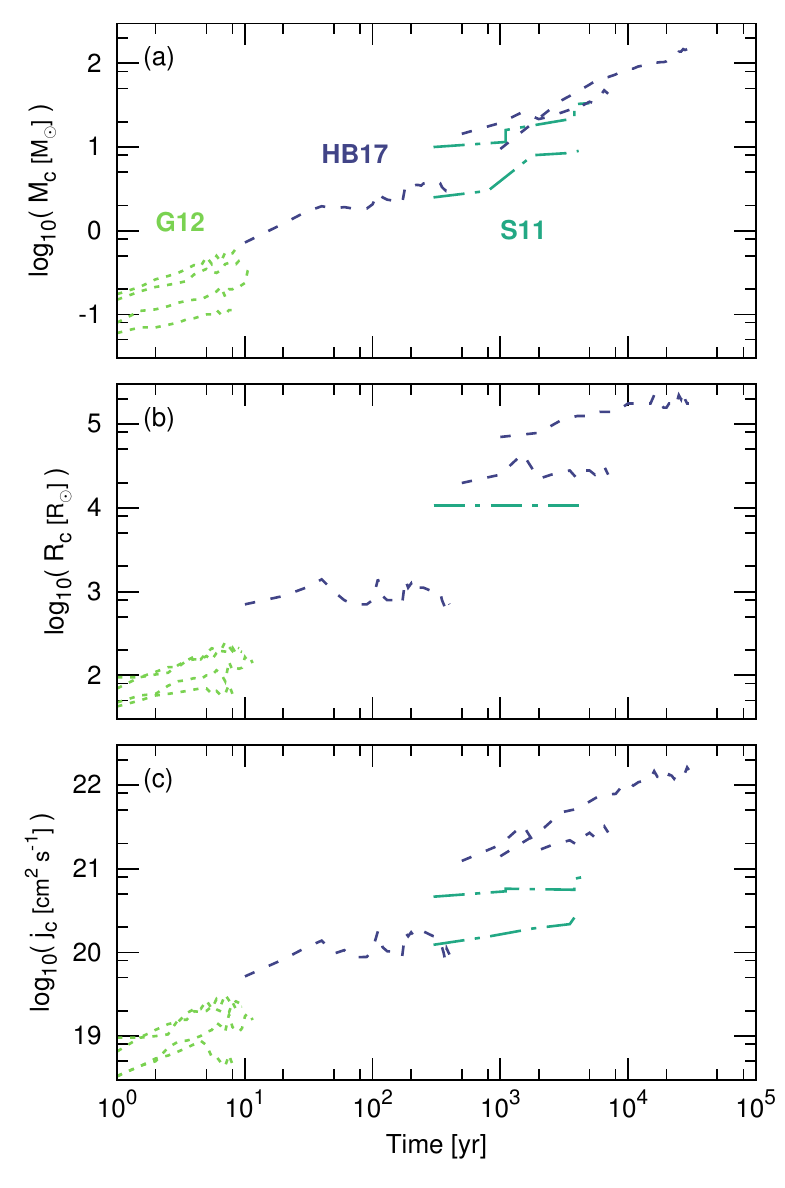}
\caption{
Summary of previous simulation data, with different numerical resolution and time coverage: ({\it a}) evolution of the hydrostatic core mass, ({\it b}) core radius, and ({\it c}) specific angular momentum.
We show data from three studies: G12 \citep[][dotted lines]{greif12,stacy13rot}, HB17 \citep[][dashed]{hirano17a}, and S11 \citep[][dot-dashed]{stacy11rot}.
None of these simulations includes magnetic fields.
}
\label{figA1}
\end{figure}

We consider previous simulation data of Pop~III star formation, extracting estimates for the resulting stellar rotation, for comparison with our modelling (Fig.~\ref{figA1}). 
\cite{greif12} is the only simulation to date which applied no sub-grid model, and directly resolved the optically thick protostellar core.
However, their time coverage is limited to only $\sim\!10$\,yr after initial protostellar core formation.
Specifically, we use results for four different haloes (MH1 to MH4). 
\cite{hirano17a} perform a series of simulations, resulting in hydrostatic cores with threshold densities $\nh = 10^{15}$, $10^{12}$ and $10^{10}\,\cc$, above which gas collapse is suppressed by artificially enhanced opacity.
These simulations were initiated from the same large-scale star forming site used in our baseline model \citep{hirano14}.
We analyse the primary hydrostatic core in these three cases.
\cite{stacy11rot} compute gas cloud collapse up to $\nh = 10^{12}\,\cc$ by adopting the sink particle method with sink radius $50\,{\rm au} \simeq 10750\,\rsun$.
We consider the growth of two cases (sinks A and B) in that study.

\bsp
\label{lastpage}
\end{document}